%% file: main.tex
\documentclass{Interspeech}

\usepackage{booktabs} % For better table formatting
\usepackage{multirow}
\usepackage{xcolor}
\usepackage{graphicx}
\usepackage{comment}
\usepackage{adjustbox}
\usepackage{caption}
\usepackage{amsmath,amssymb,amsfonts}

\interspeechcameraready

\title{G-IFT: A Gated Linear Unit adapter with Iterative Fine-Tuning for Low-Resource Children's Speaker Verification}

\author[affiliation={}]{Vishwas}{Shetty*}
\author[affiliation={}]{Jiusi}{Zheng*}
\author[affiliation={}]{Abeer}{Alwan}

\affiliation{Department of Electrical and Computer Engineering}{, University of California, Los Angeles}{USA}
\email{shettyvishwas@ucla.edu, zheng94@ucla.edu, alwan@ee.ucla.edu}


\usepackage{comment}

\begin{document}

\maketitle
\renewcommand{\thefootnote}{\fnsymbol{footnote}}
\footnotetext[1]{These authors contributed equally to this work.}
\renewcommand{\thefootnote}{\arabic{footnote}}  
\begin{abstract}
Speaker Verification (SV) systems trained on adults speech often underperform on children's SV due to the acoustic mismatch, and limited children speech data makes fine-tuning not very effective. In this paper, we propose an innovative framework, a Gated Linear Unit adapter with Iterative Fine-Tuning (G-IFT), to enhance knowledge transfer efficiency between the high-resource adults speech domain and the low-resource children's speech domain. In this framework, a Gated Linear Unit adapter is first inserted between the pre-trained speaker embedding model and the classifier. Then the classifier, adapter, and pre-trained speaker embedding model are optimized sequentially in an iterative way. This framework is agnostic to the type of the underlying architecture of the SV system. Our experiments on ECAPA-TDNN, ResNet, and X-vector architectures using the OGI and MyST datasets demonstrate that the G-IFT framework yields consistent reductions in Equal Error Rates compared to baseline methods.

% Speaker Verification (SV) systems trained using adult speech often experience challenges in performance when applied to children's SV. Due to the scarcity of children's speech data, fine-tuning SV models pre-trained on adult speech using child speech usually fails to mitigate acoustic mismatches, hindering effective knowledge transfer and domain adaptation. In this paper, we propose an innovative framework, a Gated Linear Unit adapter with Iterative Fine-Tuning (G-IFT), to enhance knowledge transfer efficiency between the high-resource adult speech domain and the low-resource children's speech domain. %thereby mitigating the domain mismatch problem in children's SV. Within the G-IFT framework, a Gated Linear Unit adapter is first inserted between the pre-trained speaker embedding model and the classifier. Then the classifier, adapter, and pre-trained speaker embedding model are optimized sequentially in an iterative manner. 
% This framework is agnostic to the type of the underlying architecture of the SV system. We evaluate the G-IFT framework on three widely used architectures: ECAPA-TDNN, ResNet, and X-vector, using two benchmark datasets, OGI and MyST. Our proposed approach consistently achieves lower Equal Error Rates compared to the baseline methods.  
\end{abstract}

\section{Introduction}
With the rise of technological innovations, children increasingly engage with social media and e-learning platforms. While these tools foster development, they also raise security concerns, necessitating protective measures. Advancements in voice-based Children's Speaker Verification (C-SV) systems are a crucial step in ensuring children's online safety \cite{graafland2018new}. 
%Nowadays, Children's Speaker Verification (C-SV) systems have garnered increasing attention due to their potential applications in safeguarding children in digital environments %such as e-learning platforms, virtual assistants, and social media 
These systems aim to accurately verify a child's identity based on their voice, addressing security concerns and ensuring personalized, safe user experiences.

Compared to other domain adaptation problems, a major challenge in C-SV arises from the constraints imposed by the scarcity of children's speech data. Although Speaker Verification (SV) systems \cite{desplanques2020ecapa, zhou2021resnext,lim2024improving} trained on adult speech have achieved notable success across various datasets \cite{nagrani2020voxceleb, fan2020cn}, they often face challenges when applied to children speech due to the significant acoustic differences between adult and children speech \cite{Child_adult_acoustic_differeces)}. Fine-tuning pre-trained models initially trained on adult speech with children speech datasets is one of the widely adopted strategies. However, fine-tuning pre-trained verification models with limited children's speech data may not adequately address the acoustic mismatch due to the scarcity of children's speech resources, resulting in inefficient knowledge transfer and hence subpar adaptation to the target domain \cite{graave2024mixed, fan22d_interspeech}. 

%Fine-tuning pre-trained models initially trained on adult speech with children speech datasets %\cite{batliner2005pf_star,cmu_kids_corpus} is one of the widely adopted strategies \cite{abed2024deep}. %However, fine-tuning pre-trained verification models with limited children speech data \cite{shahnawazuddin2021children} may not adequately address the acoustic mismatch due to the limited resources, resulting in reduced knowledge transfer efficiency and subpar adaptation to the target domain. 

To address the data scarcity challenge in C-SV, researchers have primarily adopted two strategies. The first is out-of-domain data augmentation  \cite{kathania2024spectral,aziz2024role,aziz2024short,aziz2024enhancing}, which supplement existing child speech data by adding perturbed adult speech data so that its acoustic properties resemble children's speech. This includes methods such as perturbing adult speech 
%to resemble children speech 
by modifying parameters like speaking rate, pitch, duration, and vocal tract length, as well as leveraging cycle-consistent GAN-based voice conversion to transform adult speech into child-like speech \cite{aziz2024role, aziz2024short,aziz2024enhancing}.

\begin{figure*}[t!]
    \centering
    \scalebox{1.5}{
    \begin{adjustbox}{width=1.25\columnwidth}
    \includegraphics[width=\textwidth]{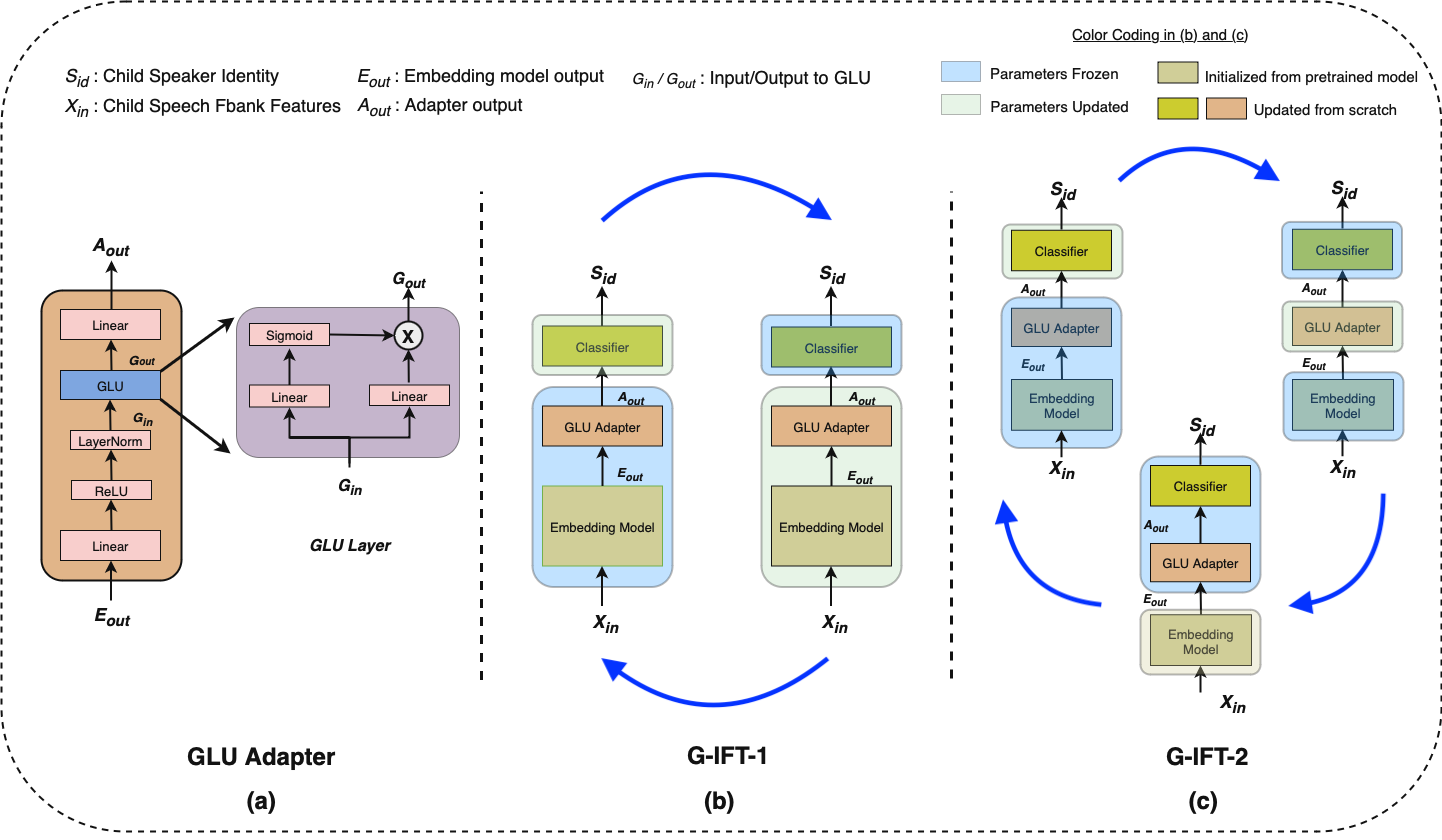} % Replace with your image file
    \end{adjustbox}}
    \caption{\textit{An overview of the proposed G-IFT framework. The GLU adapter architecture is shown in (a), and illustration of the steps involved in the G-IFT-1 and G-IFT-2 methods (discussed in Section \ref{sec:IFT}) are shown in (b) and (c) respectively.}} %for adapting a pre-trained adult speaker embedding model to child speech domain.}}
    \label{fig:G-IFT_framework}
\end{figure*}

Another strategy involves model-level adaptations, including optimizing model architectures \cite{radha2024automatic}, designing more effective loss functions \cite{kataria2020analysis}, and fine-tuning of the pre-trained adult SV model using children speech \cite{shetty2025enhancing}. %and incorporating lightweight adapters into pre-trained models. 
Adapter-based fine-tuning, achieved by integrating lightweight adapter modules into pre-trained speaker embedding models, effectively enhances the efficiency of knowledge transfer compared to the baseline methods. Although adapter-based approaches have shown promise in domain adaptation tasks \cite{sang2024efficient,huang2024robust,wang24ma_interspeech,han2023short,chen2023chapter,peng2023parameter}, to the best of our knowledge, no prior work has explored their use for C-SV, where data scarcity presents greater challenges and distinguishes it from other adaptation tasks such as cross language adaptation \cite{xia2019cross}.

% Adapter-based approaches \cite{sang2024efficient,huang2024robust,wang24ma_interspeech,han2023short,chen2023chapter,peng2023parameter}, have achieved good results for adult SV. %emerged as a promising approach in adult SV.
% %Adapter-based fine-tuning \cite{huang2024robust, sang2024efficient, peng2023parameter}, in particular, has emerged as a kind of promising approach for domain adaptation. 
% However, to the best of our knowledge, no prior work has explored the use of adapter-based frameworks for C-SV, nor anyone has proposed iterative fine-tuning for adult or child SV.%to mitigate domain mismatches in CSV.

In this paper, we propose a novel framework, Gated Linear Unit adapter (GLU) with Iterative Fine-Tuning, which we refer to as G-IFT, to enhance knowledge transfer efficiency between the adults speech domain and the children speech domain. We first propose a novel adapter module using GLU to fine-tune the speaker embedding models trained on adult speech, alleviating the domain shifting problem in C-SV. Additionally, we introduce a novel iterative fine-tuning strategy, which builds upon the GLU adapter fine tuning, wherein different components of the model are alternatively fine-tuned to enhance the SV accuracy. Experimental results indicate that our method amplifies the positive impact of adapter fine-tuning particularly in the low resource scenario, achieving a further reduction in the EER of the SV systems. The proposed G-IFT framework also exhibits potential for broader applications in other low-resource SV tasks, including verification tasks for individuals with speech disorders and similar under-resourced scenarios.

The following sections are structured as follows: Section 2 introduces the GLU adapter and iterative fine-tuning in the G-IFT framework. Section 3 covers the experimental setup. Section 4 presents the results of the proposed approach under varying conditions. Section 5 concludes and suggests future work.

\section{Methods}
% In this section, we introduce the proposed framework: Gated Linear Unit based adapter with Iterative Fine-Tuning (G-IFT). Fig.1 provides an overview of G-IFT and the structure of the GLU adapter.
\label{sec:method}
\subsection{Gated Linear Unit adapters}

We first propose a novel adapter structure inspired from the Gated Linear Unit (GLU) mechanism \cite{dauphin2017language}. The motivation behind using a GLU layer was to allow the network to decide how much information should flow through the adapter, i.e., gating weights based on the adaptation data. The GLU layer learns the gating weights  with the help of two linear layers and a sigmoid operation. The GLU adapter structure used in this paper is shown in Figure \ref{fig:G-IFT_framework} (a). Firstly, the output of the speaker embedding model, $E_\text{out}$ is fed into the adapter module, passing through a linear layer followed by a ReLU activation function. Then after performing layer normalization, $G\_in$ is passed through the GLU layer. Finally, the output of GLU layer, $G\_out$ is passed through a linear layer before being passed on to perform speaker identification by the classifier. The operation performed by GLU layer is given in Equation (\ref{eq:GLU}). $W$, $V$, $b$, and $c$ are learnable parameters. In GLU layer, the sigmoid operation on the output of one of the linear layers acts as the weights to the output of the second linear layer, deciding how much information from the pre-trained embedding model should be passed on ahead.

\begin{equation}
    G_{out} = (G_{in} \ast W + b) \otimes \sigma(G_{in} \ast V + c)
    \label{eq:GLU}
\end{equation}

The proposed adapter is inserted between the output of the speaker embedding model and the input of the classifier. When evaluating the proposed GLU adapter based approach, the outputs of the adapter module, denoted as $A_{\text{out}}$ in Figure \ref{fig:G-IFT_framework}, are utilized as speaker embeddings. %This adapter module can be used across different embedding model architectures. 

%During test time, given two speakers audio segments. the output of the adapter module $A_{out}$ is used as the speaker embeddings. This adapter module can be used across different embedding model architectures. 

% In this work, we evaluate the proposed adapter framework across three speaker verification architectures: ECAPA-TDNN \cite{desplanques20_interspeech}, ResNet \cite{he2016deep}, and Xvector \cite{snyder2018x}-based systems.
%In this work, our focus has been to propose an architecture that gives improvements over the base adapter while not adding extra parameter overhead. 

%\subsection{Iterative Fine-Tuning Strategy in G-IFT Framework}
\subsection{G-IFT: GLU adapter with Iterative Fine-Tuning}
\label{sec:IFT}
%This section aims to 
Fine-tuning pretrained verification models in SV tasks differs from fine-tuning in other tasks such as Automatic Speech Recognition (ASR) due to the need for adjusting the classifiers based on the number of speakers in the target domain dataset. This requires the classifier’s parameters to be trained from scratch, rather than relying solely on transfer learning. Traditional adapter fine-tuning methods commonly update the adapter and classifier simultaneously; in some cases, the adapter, classifier, and embedding model are all updated simultaneously. However, since the classifier and the inserted adapters typically require more substantial adjustments than the pretrained embedding model, this approach can be inefficient especially in the low-resource scenario, potentially slowing down convergence and reducing overall performance.

To enhance the efficiency of knowledge transfer, we propose a novel framework, Gated Linear Unit Adapter with Iterative Fine-Tuning (G-IFT), which applies an iterative fine-tuning strategy to the GLU adapter. The motivation behind this approach is to address potential inefficiencies associated with simultaneous updates by allowing more focused updates to the adapter and classifier initially, which helps them adapt more effectively to in-domain children speech data. This framework aims to make more efficient use of target data during fine-tuning, especially in the low-resource scenario, which leads to better performance in C-SV. Additionally, this method is highly adaptable and is not constrained by the underlying model architecture.
% \vspace{-2em}

%\textcolor{red}{Combining the novel adapter architecture and iterative fine-tuning methodology, we propose the Gated Linear Unit adapter - with Iterative Fine-Tuning (G-IFT) framework.}
%G-IFT is designed to alleviate the domain shifting problem for C-SV. 
The proposed G-IFT framework includes two variants, G-IFT-1 and G-IFT-2, which share the same foundational structure but differ in their iterative fine-tuning strategies as shown in Figure \ref{fig:G-IFT_framework} (b) and (c). In G-IFT-1, the process involves three main steps. First, the model is initialized and trained on the source domain dataset to obtain a pre-trained speaker embedding model. Second, the GLU adapter is inserted between the pretrained speaker embedding model and a classifier designed for fine-tuning on the target domain dataset. Third, the training alternates iteratively between fine-tuning the GLU adapters and the classifier jointly, followed by fine-tuning the pretrained speaker embedding model. G-IFT-2 follows the same first two steps as G-IFT-1 but introduces a slightly different iterative strategy. In this variant, the process alternates iteratively by first fine-tuning the classifier, then the GLU adapters, and then the pretrained speaker embedding model. This adjustment allows G-IFT-2 to prioritize classifier adaptation in isolation before integrating adjustments from the adapters and embedding model.

\section{Experimental Settings}
\label{sec:experiment}
\subsection{Databases}
\label{sec:database}
We conduct experiments on two children's speech databases, OGI \cite{shobaki2000ogi} and MyST \cite{ward2011my}. For OGI, we use the scripted portion, which contains children's speech from approximately 100 speakers per grade, spanning kindergarten to grade 10. The train/eval split follows the protocol established in \cite{singh2024childaugment}.
The MyST dataset consists of 499 hours of speech data collected from 1,372 students in grades 3 to 5. For our experiments we utilize only the annotated portion totaling approximately 268 hours. For MyST, we also create multiple smaller training subsets from the MyST training set, maintaining a consistent evaluation split to analyze the impact of in-domain training data size on the proposed method. While creating the MyST train splits MyST-1 to MyST-4, we ensure that data from all the speakers in the complete training set is included. Hence, as we go from MyST-1 to MyST-4, the amount of speech data per speaker increases. Across all train and evaluation splits in both datasets, there is no overlap in speakers. It should be noted that the MyST evaluation set is the same, irrespective of the MyST train split. The train eval splits of both databases are shown in Table \ref{tab:database_details}.

%To-Do: Describe the database/scripted one

% We report results on two child speech databases, OGI \cite{shobaki2000ogi} and MyST \cite{ward2011my}. Table \ref{tab:database_details} lists the details of the train/eval splits created from these two databases. In OGI dataset, we use the train/eval split from \cite{singh2024childaugment}. For MyST, we use the train/eval splits from \cite{fan2024benchmarking}. From the MyST train split, we create multiple smaller train sets. The details of the different train splits are given in Table \ref{tab:database_details}. While creating the MyST train splits, we ensure that data from all the speakers in the complete train set is included. Hence, as the duration of the MyST train split increases, the amount of speech data per speaker increases. From the MyST evaluation split, we extract 100000 enrollment-test pairs to create the evaluation set.

\begin{table}[t!]
\centering
\caption{\textit{The train-eval splits of the OGI and MyST databases used for training/fine-tuning the models are presented. \#Spks refers to the number of speakers, \#Hrs to the duration in hours, and \#Trials to the number of SV trials in the evaluation set.}}
\resizebox{0.9\columnwidth}{!}{%
\begin{tabular}{llllc}
\hline 
\multirow{2}{*}{} & \multicolumn{2}{c}{Train} & \multicolumn{2}{c}{Eval} \\ \cline{2-3} \cline{4-5} 
                  & \# Spks  & \# Hrs  & \# Spks  & \# Trials  \\ \hline
\textit{OGI}   & 120      & 2.65     & 993      & 190972     \\
\textit{MyST-1}   & 1210      & 2.00     & 91      & 100000     \\
\textit{MyST-2}   & 1210      & 8.00     & 91      & 100000     \\
\textit{MyST-3}   & 1210      & 85.00     & 91      & 100000     \\
\textit{MyST-4}   & 1210     & 268.00     & 91      & 100000     \\\hline
\end{tabular}%
}
\label{tab:database_details}
\end{table}

%\subsection{Speaker Verification Systems}
\subsection{Baseline Systems}
\label{sec:baseline}
In this study, we utilized ECAPA-TDNN \cite{desplanques2020ecapa}, ResNet \cite{VILLALBA2020101026}, and X-vector \cite{snyder2018spoken} as the SV systems. All experiments were conducted using the SpeechBrain toolkit \cite{ravanelli2021speechbrain}. Models trained from scratch on the OGI and MyST datasets served as the baseline systems and are referred to as \textit{Baseline}. Pretrained speaker embedding models are trained on the VoxCeleb dataset \cite{nagrani2020voxceleb}. The experiment where these pretrained models are directly tested on children speech data is referred to as \textit{Pretrained}. We refer to fine-tuning the  \textit{Pretrained} model with child speech data for a fixed number of epochs as vanilla fine-tuning and this model is referred to as \textit{Finetune} in this paper. The input features for all the models were 80-dimensional filter bank features extracted with a frame length of 25 ms and a hop size of 10 ms. Equal Error Rate was used to evaluate the SV systems' performance.

%The pretrained models are fine-tuned using MyST and OGI speech datasets.

%During fine-tuning, we employed the Additive Angular Margin (AAM) loss function for ECAPA-TDNN and ResNet and the negative log-likelihood loss function for X-Vector, consistent with the loss functions used during their respective pre-training phases.

%To ensure a fair comparison, we maintained consistency in key training configurations, such as the optimizer, learning rate, and loss function, across all methods which used the same speaker verification systems and datasets. 

Optimization was performed using the Adam optimizer with a learning rate of 0.001 and a weight decay of 0.000002, along with a cyclic learning rate scheduler to balance exploration and exploitation, with a base learning rate of $1 \times 10^{-8}$, a maximum learning rate of 0.001, and a step size of 65,000. Notably, the number of training epochs varied across different methods due to differences in fine-tuning strategies. Specifically, the vanilla fine-tuning method updates all model parameters, including both the speaker embedding model and classifier, simultaneously within each training epoch. In contrast, the G-IFT framework adopts an iterative strategy, alternately fine-tuning the classifier, GLU adapters, and the pretrained speaker embedding model in a iterative manner. As a result, the total number of epochs for the GIFT-1 method was twice that of the vanilla fine-tuning method, and for the GIFT-2 method, it was three times that of the vanilla fine-tuning method. This adjustment ensures that the number of parameter updates across all methods is equivalent.

% \subsection{Proposed G-IFT framework}
% \label{sec:gift_framework}
% The G-IFT framework adopts an iterative strategy, alternately fine-tuning the classifier, GLU Adapters, and the pretrained speaker embedding model in a cyclic manner. As a result, the total number of epochs for the GIFT-1 method was twice that of the vanilla fine-tuning method, and for the GIFT-2 method, it was three times that of the vanilla fine-tuning method. This adjustment ensures that the number of parameter updates across all methods is equivalent.

\section{Results and Discussion}
\label{sec:results}
\renewcommand{\arraystretch}{0.95}
\begin{table}[t]
    \centering
    \caption{\textit{Performance in terms of Equal Error Rate (EER \%) comparison across OGI and MyST-1 datasets. \textbf{Pretrained} is the open source adult SV systems tested directly on children speech data. \textbf{Baseline} is the SV systems trained using children speech data from scratch, and \textbf{Finetuned} is the adult SV systems finetuned using children speech data. \textbf{GLU} is adapter fine tuning, \textbf{G-IFT-1}, and \textbf{G-IFT-2} are the proposed adapter fine tuning framework. \textbf{RA} is the Residual Adapter \cite{fan22d_interspeech}}. Significant improvements (paired t-test; p-value=0.05) over Finetune are represented with ``*''.}
    {\Large
    \resizebox{0.48\textwidth}{!}{%
    \begin{tabular}{lcc|cc|cc}
        \toprule
        & \multicolumn{2}{c|}{\textbf{ECAPA-TDNN}} & \multicolumn{2}{c|}{\textbf{ResNet}} & \multicolumn{2}{c}{\textbf{X-vector}} \\
        \cmidrule(r){2-3} \cmidrule(r){4-5} \cmidrule(r){6-7}
        & OGI & MyST-1 & OGI & MyST-1 & OGI & MyST-1 \\
        \midrule
        Pretrained & 17.39  & 17.48  & 12.62  & 14.47  & 23.59  & 24.95 \\
        Baseline   & 12.15  & 24.80  & 14.23  & 36.23  & 14.71  & 25.70  \\
        Finetune   & 11.10  & 20.04  & 7.45  & 10.64  & 15.89   & 21.26 \\
        RA \cite{fan22d_interspeech}   & 11.34  & 22.89  & 7.47  & 12.76  & 15.61   & 21.54 \\
        GLU        & 9.70   & 22.30  & 6.92  & 12.48  & 12.31   & 17.96  \\
        G-IFT-1     &  \hphantom{0}\textbf{8.88*}   & 16.38  & \hphantom{0}\textbf{6.77*}   & 9.29  & 12.71   & 18.20 \\
        G-IFT-2     & 9.06   & \hphantom{0}\textbf{14.22*}  & 6.88   & \hphantom{0}\textbf{8.37*}  & \hphantom{0}\textbf{12.09*}   & \hphantom{0}\textbf{16.79*}  \\
        \bottomrule
    \end{tabular}}}
    \label{tab:ogi_myst_comparison}
\end{table}

\renewcommand{\arraystretch}{1.0} 

\begin{table*}[t!]
    \centering
    \caption{\small \textit{Equal Error Rate (EER \%) across different models—ECAPA-TDNN, X-vector, and ResNet—on various MyST training splits. The duration of the training splits increases progressively from MyST-1 to MyST-4. The evaluation results are obtained using a consistent testing split, with no overlap in speakers between the training and test splits in all experiments. Significant improvements (paired t-test; p-value=0.05) over Finetune are represented with ``*''.}}
    {\Large
    \resizebox{1.0\textwidth}{!}{%
    \begin{tabular}{lcccc|cccc|cccc}
        \toprule
        & \multicolumn{4}{c|}{\textbf{ECAPA-TDNN}} & \multicolumn{4}{c|}{\textbf{ResNet}} & \multicolumn{4}{c}{\textbf{X-vector}} \\
        \cmidrule(r){2-5} \cmidrule(r){6-9} \cmidrule(r){10-13}
        & MyST-1 & MyST-2 & MyST-3 & MyST-4 & MyST-1 & MyST-2 & MyST-3 & MyST-4 & MyST-1 & MyST-2 & MyST-3 & MyST-4 \\
        \midrule
        Baseline & 24.80   & 14.42  & 10.76  & 10.11  & 36.23  & 19.26  & 18.05  & 11.48  & 25.70   & 19.76 & 17.08 & 16.24 \\
        Finetune & 20.04   & 7.97   & 5.81   & 6.05   & 10.64  & 9.33  & 6.60    & 4.87   & 21.26  & 19.16  & 16.70   & 16.11  \\
        RA \cite{fan22d_interspeech}  & 22.89 & 8.19   & 6.13  & 6.07   & 12.76   & 8.62   & 5.70    & 4.87   & 21.54  & 19.50  & 16.36 & 17.83  \\
        GLU      & 22.30 & 7.51   & 5.92  & 5.97   & 12.48   & 7.48   & 5.50    & 5.07   & 17.96  & 15.05  & \hphantom{0}\textbf{11.49*}  & \hphantom{0}\textbf{11.75*}  \\
        G-IFT-1   & 16.38 & \textbf{7.03}   & \hphantom{0}\textbf{5.49*}  & \hphantom{0}\textbf{5.42*}   & 9.29   & 6.42   & \hphantom{0}\textbf{4.89*}   & \textbf{4.82}   & 18.20   & 15.37  & 12.29  & 12.58  \\
        G-IFT-2   & \hphantom{0}\textbf{14.22*} & 7.81   & 5.93   & 5.99   & \hphantom{0}\textbf{8.37*}   & \hphantom{0}\textbf{5.87*}   & 5.52   & 5.27   & \hphantom{0}\textbf{16.79*}  & \hphantom{0}\textbf{14.58*}  & 14.70   & 15.30   \\
        \bottomrule
    \end{tabular}}}
    \label{tab:myst_performance}
\end{table*}

\subsection{Performance of the G-IFT framework}
\label{sec:ogi_myst_comparison}
Table~\ref{tab:ogi_myst_comparison} presents results on the OGI and MyST-1 datasets using various training methods. The proposed G-IFT framework, including G-IFT-1 and G-IFT-2, consistently improves performance across all architectures. Rows 2 and 3 show baseline and vanilla fine-tuning results. Fine-tuning generally outperforms training from scratch, except for X-vector on OGI. Row 4 reports results of Residual Adapter (RA) fine-tuning, configured to match our GLU adapter in parameter size and training steps (15 epochs). Although RA performed well in \cite{fan22d_interspeech}, it is less effective here. In contrast, direct fine-tuning with the GLU adapter (without iteration) outperforms vanilla fine-tuning in four of six cases. The last two rows show that G-IFT-1 and G-IFT-2 consistently outperform vanilla fine-tuning in all test cases. Compared to GLU fine-tuning, one of the G-IFT variants always achieves the best result, validating the benefit of iterative strategies in enhancing GLU adapter adaptability under limited data. We also examined Iterative Fine Tuning (IFT) alone on MyST-1, which yields EERs of 22.5\%, 10.2\%, and 20.9\% for ECAPA-TDNN, ResNet, and x-vector, without clear improvement. In contrast, applying IFT to GLU reduces EERs to 14.2\%, 8.4\%, and 16.8\%, The adapter-classifier setup provides greater modeling capacity.

\subsection{Impact of Training Data Size on Model Adaptability}
To further investigate the impact of training dataset size on the proposed G-IFT framework, we use the four MyST subsets (Table \ref{tab:database_details}) to analyze the performance as training resources increased. As shown in Table \ref{tab:myst_performance}, the EERs of all methods decrease overall with increasing training data. Comparing \textit{Finetune} with \textit{Baseline} in Table \ref{tab:myst_performance}, ECAPA-TDNN and ResNet exhibit significantly better performance with fine-tuning using larger MyST datasets, while X-vector shows comparable results between the two approaches. This discrepancy can likely be attributed to differences in model complexity. The larger and more intricate architectures of ECAPA-TDNN and ResNet may better fit the children speech data when trained using larger amounts of in-domain data, whereas the simpler architecture of X-vector might not.
Residual Adapter (RA) fine-tuning results are given in row 3 of Table \ref{tab:myst_performance}. Additionally, GLU adapter i.e., row 4 of Table \ref{tab:myst_performance} performs better than RA in 11 out of 12 test cases. 

\begin{figure}[t]
    \centering
    \scalebox{1.0}{
        \includegraphics[width=1.0\columnwidth]{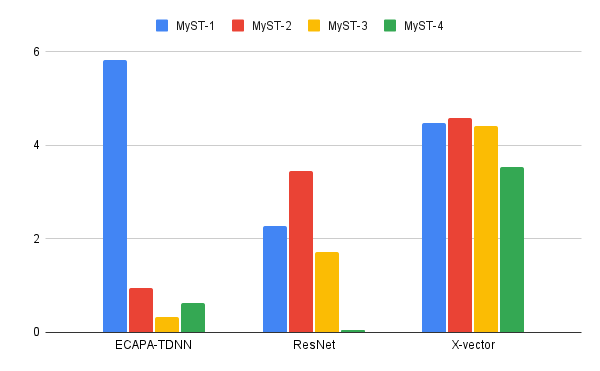} % Replace with your image file
    }
    \caption{\textit{Absolute reduction in EER (\%) (y-axis) by the proposed G-IFT framework (Rows 4–5, Table \ref{tab:myst_performance}) over Finetune (Row 2) on different MyST training splits. Larger reductions are observed in low-resource settings (MyST-1/2) than in high-resource ones (MyST-3/4), underscoring the framework’s effectiveness in low-resource scenarios.}} %The improvements achieved are statistically significant (p-value=0.05).}}
    \label{fig:rel_imp}
    \vspace{-10pt}
\end{figure}

In terms of the G-IFT framework, all three model architectures achieve lower EERs across different MyST training subsets compared to vanilla fine-tuning. Specifically, as shown in Figure~\ref{fig:rel_imp}, the absolute reduction in EER (\%) obtained by G-IFT-1 and G-IFT-2 methods on the MyST-1 and MyST-2 datasets are more pronounced and consistent than those on MyST-3 and MyST-4 datasets across three architectures, highlighting the effectiveness of our proposed approach particularly in the low-resource scenarios. This phenomenon is likely due to our method’s focus on iteratively fine-tuning the adapter and classifier first, which helps mitigate embedding overfitting.  However, when sufficient training data is available, the large dataset naturally reduces embedding overfitting, making the efficiency of our method less critical and diminishing its advantage.
We also observed that both G-IFT-1 and G-IFT-2 outperform vanilla fine-tuning on MyST-1 and MyST-2 across all three model architectures, though the best-performing method varied between the  G-IFT-1 and G-IFT-2. However, with larger training datasets like MyST-3 and MyST-4, G-IFT-1 consistently achieved better results than G-IFT-2. A possible explanation is that in low-resource scenarios, sequentially fine-tuning individual components - the classifier , adapter and base model, as implemented in G-IFT-2 may be more effective. Conversely, under high-resource conditions, it appears sufficient to update relatively larger sub-modules, as in G-IFT-1, where the adapter and classifier are updated jointly followed by base model update. This hypothesis is further supported by the observation that vanilla fine-tuning also performs well in high-resource settings, as reflected in our results in Table \ref{tab:myst_performance}. These observations prompt us to further investigate the respective strengths of G-IFT-1 and G-IFT-2 in future work, in order to refine the framework and propose a more universally effective approach applicable across diverse scenarios.

\section{Conclusion}
\label{sec:conclusion}

%\vspace{-0.25em}

%In this letter, we proposed the G-IFT framework designed to alleviate the domain mismatch problem in Children Speaker Verification under the low-resource scenario. 
% In this paper, we proposed the G-IFT 
% framework to address the domain mismatch challenge when adapting adult, i.e, high resource, SV models  to perform child, i.e, low resource, SV.
%G-IFT a framework integrating a GLU based adapter and 
In this paper, we proposed the G-IFT framework to alleviate the domain mismatch challenge when adapting adults (i.e., high-resource) SV models to perform children (i.e., low-resource) SV. Through the integration of the GLU adapter and %the implementation of along with 
an iterative fine-tuning strategy, the proposed framework achieved consistent performance improvements across three neural network architectures on the OGI and MyST datasets. The G-IFT framework outperforms baseline methods, achieving competitive performance %with
using significantly %reduced 
lesser in-domain data. %We also conduct ablation studies using 
By varying amounts of in-domain training data using MyST dataset we evaluate the framework's performance under different resource constraints and our results highlight the potential of the proposed G-IFT framework in improving speaker verification systems in low-resource scenarios. 
Future work will focus on refining the G-IFT framework and extending its application to other low-resource SV tasks, including those involving disordered speech.

%We also conduct ablation studies using varying amounts of in-domain data to evaluate the framework's performance under different resource constraints.

%This paper introduces a novel framework for adapting Adult SV models to Child SV, evaluated on three neural network architectures and two child speech databases. The proposed adapter and fine-tuning approach consistently outperform baseline methods across all architectures and databases. Compared to the widely-used Residual Adapter (RA), the GLU-based adapter achieves better performance in all scenarios. Figure \ref{fig:rel_imp} shows that, except for ResNet MyST-4, our approach (GIFT-1 or GIFT-2) delivers at least 5\% relative improvement over vanilla fine-tuning. Future work includes exploring the G-IFT framework for other tasks and ensuring pretrained models retain adult speech knowledge post-adaptation.

% A conclusion section is not required. Although a conclusion may review the main points of the paper, do not replicate the abstract as the conclusion. A conclusion might elaborate on the importance of the work or suggest applications and extensions. 

\section{Acknowledgment}
The research is supported in part by the NSF and the IES, U.S. Department of Education (DoE), through Grant R305C240046 to the U. at Buffalo. The opinions expressed are those of the authors and do not represent views of the IES,  DoE, or the NSF.

\newpage
\bibliographystyle{IEEEtran}
%\bibliography{main}
\input{main.bbl}

\end{document}

%% file: main.bbl

%% file: main.bbl
\begin{thebibliography}{10}
\providecommand{\url}[1]{#1}
\csname url@samestyle\endcsname
\providecommand{\newblock}{\relax}
\providecommand{\bibinfo}[2]{#2}
\providecommand{\BIBentrySTDinterwordspacing}{\spaceskip=0pt\relax}
\providecommand{\BIBentryALTinterwordstretchfactor}{4}
\providecommand{\BIBentryALTinterwordspacing}{\spaceskip=\fontdimen2\font plus
\BIBentryALTinterwordstretchfactor\fontdimen3\font minus \fontdimen4\font\relax}
\providecommand{\BIBforeignlanguage}[2]{{%
\expandafter\ifx\csname l@#1\endcsname\relax
\typeout{** WARNING: IEEEtran.bst: No hyphenation pattern has been}%
\typeout{** loaded for the language `#1'. Using the pattern for}%
\typeout{** the default language instead.}%
\else
\language=\csname l@#1\endcsname
\fi
#2}}
\providecommand{\BIBdecl}{\relax}
\BIBdecl

\bibitem{graafland2018new}
J.~H. Graafland, ``New technologies and 21st century children: Recent trends and outcomes,'' \emph{Organization for Economic Co-operation and Development Working Papers (OECD)}, 2018.

\bibitem{desplanques2020ecapa}
B.~Desplanques, J.~Thienpondt, and K.~Demuynck, ``{ECAPA-TDNN}: Emphasized {Channel} {Attention}, {Propagation} and {Aggregation} in {TDNN Based} {Speaker Verification},'' in \emph{Interspeech}, 2020, pp. 3830--3834.

\bibitem{zhou2021resnext}
T.~Zhou, Y.~Zhao, and J.~Wu, ``Resnext and res2net structures for speaker verification,'' in \emph{IEEE Spoken Language Technology Workshop (SLT)}, 2021, pp. 301--307.

\bibitem{lim2024improving}
C.~yeong Lim, H.~seo Shin, J.~ho~Kim, J.~Heo, K.-W. Koo, S.~bin Kim, and H.-J. Yu, ``Improving noise robustness in self-supervised pre-trained model for speaker verification,'' in \emph{Interspeech}, 2024, pp. 2665--2669.

\bibitem{nagrani2020voxceleb}
A.~Nagrani, J.~S. Chung, W.~Xie, and A.~Zisserman, ``Voxceleb: Large-scale speaker verification in the wild,'' \emph{Computer Speech \& Language}, vol.~60, p. 101027, 2020.

\bibitem{fan2020cn}
Y.~Fan, J.~Kang, L.~Li, K.~Li, H.~Chen, S.~Cheng, P.~Zhang, Z.~Zhou, Y.~Cai, and D.~Wang, ``{CN-Celeb}: A challenging chinese speaker recognition dataset,'' in \emph{IEEE International Conference on Acoustics, Speech and Signal Processing (ICASSP)}, 2020, pp. 7604--7608.

\bibitem{Child_adult_acoustic_differeces)}
E.~T. Stathopoulos, J.~E. Huber, and J.~E. Sussman, ``Changes in acoustic characteristics of the voice across the life span: Measures from individuals 4–93 years of age,'' \emph{Journal of Speech, Language, and Hearing Research}, vol.~54, no.~4, pp. 1011--1021, 2011.

\bibitem{graave2024mixed}
T.~Graave, Z.~Li, T.~Lohrenz, and T.~Fingscheidt, ``Mixed children/adult/childrenized fine-tuning for children’s asr: How to reduce age mismatch and speaking style mismatch,'' in \emph{Interspeech}, 2024, pp. 5188--5192.

\bibitem{fan22d_interspeech}
R.~Fan and A.~Alwan, ``Draft: A novel framework to reduce domain shifting in self-supervised learning and its application to children’s asr,'' in \emph{Interspeech}, 2022, pp. 4900--4904.

\bibitem{kathania2024spectral}
H.~K. Kathania, V.~Kadyan, S.~R. Kadiri, and M.~Kurimo, ``Spectral warping based data augmentation for low resource children’s speaker verification,'' \emph{Multimedia Tools and Applications}, vol.~83, no.~16, pp. 48\,895--48\,906, 2024.

\bibitem{aziz2024role}
S.~Aziz and S.~Shahnawazuddin, ``Role of data augmentation and effective conservation of high-frequency contents in the context children’s speaker verification system,'' \emph{Circuits, Systems, and Signal Processing}, vol.~43, no.~5, pp. 3139--3159, 2024.

\bibitem{aziz2024short}
S.~Aziz, Ankita, and S.~Shahnawazuddin, ``Short-utterance-based children’s speaker verification in low-resource conditions,'' \emph{Circuits, Systems, and Signal Processing}, vol.~43, no.~3, pp. 1715--1740, 2024.

\bibitem{aziz2024enhancing}
S.~Aziz and S.~Shahnawazuddin, ``Enhancing children’s short utterance-based asv using inverse gamma-tone filtered cepstral coefficients,'' \emph{Circuits, Systems, and Signal Processing}, vol.~43, no.~5, pp. 3020--3041, 2024.

\bibitem{radha2024automatic}
K.~Radha, M.~Bansal, and R.~B. Pachori, ``Automatic speaker and age identification of children from raw speech using sincnet over erb scale,'' \emph{Speech Communication}, vol. 159, p. 103069, 2024.

\bibitem{kataria2020analysis}
S.~Kataria, P.~S. Nidadavolu, J.~Villalba, and N.~Dehak, ``Analysis of deep feature loss based enhancement for speaker verification,'' in \emph{The Speaker and Language Recognition Workshop (Odyssey)}, 2020, pp. 459--466.

\bibitem{shetty2025enhancing}
V.~M. Shetty, J.~Zheng, S.~M. Lulich, and A.~Alwan, ``Enhancing age-related robustness in children speaker verification,'' in \emph{ICASSP 2025-2025 IEEE International Conference on Acoustics, Speech and Signal Processing (ICASSP)}.\hskip 1em plus 0.5em minus 0.4em\relax IEEE, 2025, pp. 1--5.

\bibitem{sang2024efficient}
M.~Sang and J.~H. Hansen, ``Efficient adapter tuning of pre-trained speech models for automatic speaker verification,'' in \emph{IEEE International Conference on Acoustics, Speech and Signal Processing (ICASSP)}, 2024, pp. 12\,131--12\,135.

\bibitem{huang2024robust}
W.~Huang, B.~Han, S.~Wang, Z.~Chen, and Y.~Qian, ``Robust cross-domain speaker verification with multi-level domain adapters,'' in \emph{IEEE International Conference on Acoustics, Speech and Signal Processing (ICASSP)}, 2024, pp. 11\,781--11\,785.

\bibitem{wang24ma_interspeech}
T.~Wang, L.~Li, and D.~Wang, ``{SE/BN Adapter}: Parametric efficient domain adaptation for speaker recognition,'' in \emph{Interspeech}, 2024, pp. 2145--2149.

\bibitem{han2023short}
S.~Han, Y.~Ahn, K.~Kang, and J.~W. Shin, ``Short-segment speaker verification using ecapa-tdnn with multi-resolution encoder,'' in \emph{IEEE International Conference on Acoustics, Speech and Signal Processing (ICASSP)}, 2023, pp. 1--5.

\bibitem{chen2023chapter}
Z.-C. Chen, Y.-S. Sung, and H.-y. Lee, ``Chapter: Exploiting convolutional neural network adapters for self-supervised speech models,'' in \emph{IEEE International Conference on Acoustics, Speech, and Signal Processing Workshops (ICASSPW)}, 2023, pp. 1--5.

\bibitem{peng2023parameter}
J.~Peng, T.~Stafylakis, R.~Gu, O.~Plchot, L.~Mo{\v{s}}ner, L.~Burget, and J.~{\v{C}}ernock{\`y}, ``Parameter-efficient transfer learning of pre-trained transformer models for speaker verification using adapters,'' in \emph{IEEE International Conference on Acoustics, Speech and Signal Processing (ICASSP)}, 2023, pp. 1--5.

\bibitem{xia2019cross}
W.~Xia, J.~Huang, and J.~H. Hansen, ``Cross-lingual text-independent speaker verification using unsupervised adversarial discriminative domain adaptation,'' in \emph{ICASSP 2019-2019 IEEE International Conference on Acoustics, Speech and Signal Processing (ICASSP)}.\hskip 1em plus 0.5em minus 0.4em\relax IEEE, 2019, pp. 5816--5820.

\bibitem{dauphin2017language}
Y.~N. Dauphin, A.~Fan, M.~Auli, and D.~Grangier, ``Language modeling with gated convolutional networks,'' in \emph{International conference on machine learning}.\hskip 1em plus 0.5em minus 0.4em\relax PMLR, 2017, pp. 933--941.

\bibitem{shobaki2000ogi}
K.~Shobaki, J.-P. Hosom, and R.~Cole, ``The {OGI} kids’ speech corpus and recognizers,'' in \emph{Proc. of ICSLP}.\hskip 1em plus 0.5em minus 0.4em\relax Citeseer, 2000, pp. 564--567.

\bibitem{ward2011my}
W.~Ward, R.~Cole, D.~Bolanos, C.~Buchenroth-Martin, E.~Svirsky, S.~V. Vuuren, T.~Weston, J.~Zheng, and L.~Becker, ``My science tutor: A conversational multimedia virtual tutor for elementary school science,'' \emph{ACM Transactions on Speech and Language Processing (TSLP)}, vol.~7, no.~4, pp. 1--29, 2011.

\bibitem{singh2024childaugment}
V.~P. Singh, M.~Sahidullah, and T.~Kinnunen, ``Childaugment: Data augmentation methods for zero-resource children's speaker verification,'' \emph{The Journal of the Acoustical Society of America}, vol. 155, no.~3, pp. 2221--2232, 2024.

\bibitem{VILLALBA2020101026}
J.~Villalba, N.~Chen, D.~Snyder, D.~Garcia-Romero, A.~McCree, G.~Sell, J.~Borgstrom, L.~P. García-Perera, F.~Richardson, R.~Dehak, P.~A. Torres-Carrasquillo, and N.~Dehak, ``State-of-the-art speaker recognition with neural network embeddings in {NIST} {SRE18} and {Speakers} in the {Wild} evaluations,'' \emph{Computer Speech \& Language}, vol.~60, p. 101026, 2020.

\bibitem{snyder2018spoken}
D.~Snyder, D.~Garcia-Romero, A.~McCree, G.~Sell, D.~Povey, and S.~Khudanpur, ``Spoken language recognition using {X-vectors},'' in \emph{The Speaker and Language Recognition Workshop (Odyssey)}, 2018, pp. 105--111.

\bibitem{ravanelli2021speechbrain}
M.~Ravanelli, T.~Parcollet, P.~Plantinga, A.~Rouhe, S.~Cornell, L.~Lugosch, C.~Subakan, N.~Dawalatabad, A.~Heba, J.~Zhong \emph{et~al.}, ``Speechbrain: A general-purpose speech toolkit,'' \emph{arXiv preprint arXiv:2106.04624}, 2021.

\end{thebibliography}
